# Arc statistics with realistic cluster potentials

# III. A systematic effect on cluster mass estimates

Matthias Bartelmann

Harvard College Observatory, MS 51, 60 Garden Street, Cambridge, MA 02138, USA, and Max-Planck-Institut für Astrophysik, Postfach 1523, D–85740 Garching, FRG

20 October 1994

**Abstract.** It is shown in this paper that deviations of galaxy cluster lenses from spherical symmetry can render mass estimates for galaxy clusters based on the formation of large arcs systematically to high. Numerical models show that the mass needed for producing large arcs in clusters can be notably smaller than expected from simple spherically symmetric lens models. The reason is that the enhanced tidal effect in asymmetric and substructured lenses can compensate for part of the convergence necessary for strong lensing effects. An analytic argument is given to explain why deviations from radial symmetry will in general decrease the required lens mass. Mass estimates assuming radially symmetric lenses are on average too high by a factor of $\simeq 1.6$, and with a probability of $\simeq 20\%$ by a factor of $\simeq 2$.

## 1 Introduction

The mass estimates for galaxy clusters tend to become more and more controversial the closer clusters are inspected with different methods. From the velocity dispersion of cluster galaxies, an estimate for the cluster mass is readily found employing the virial theorem, but the virial mass found this way is necessarily based on the assumption that the cluster is at least close to virialization (Binney & Tremaine 1987). This need not be the case, because the relaxation timescale for cluster galaxies is on the order of the Hubble time at least if the cosmological density parameter $\Omega_0$ is close to unity.

The X-ray emission of galaxy clusters is sensitive to the depth of the clusters' gravitational potential well, hence the total X-ray emission can also be used for an estimate of the cluster mass. Such estimates usually assume a spherically symmetric gas distribution in hydrostatic equilibrium (e.g., Sarazin 1986). It appears unclear whether these assumptions are fulfilled in galaxy clusters. Significant uncertainties can arise, for instance, if the intracluster gas is multi-phased (Thomas, Fabian, & Nulsen 1987), if there are non-thermal pressure components like magnetic fields or small-scale turbulent motions (Miralda-Escudé & Babul 1994, Loeb & Mao 1994), or if recent mergers of cluster subclumps shock-heat the gas and thereby cause inhomogeneities in the gas density and temperature.

The weak lensing effect by clusters, which causes coherent distortions of the images of background galaxies (Tyson, Valdes, & Wenk 1990), can in principle be employed to reconstruct the surface-mass density of the clusters (Kaiser & Squires 1993). Applications of this method to real clusters have resulted in cluster masses which are two to three times larger than the virial mass for the cluster MS1224+20 (Fahlman et al. 1994, Carlberg et al. 1994). For the hottest and most luminous cluster in the sky however, A 2169, the reconstruction technique by Kaiser & Squires yields a mass map which is consistent with there being no mass at all (Squires 1994).

Still, the three methods for estimating gas masses mentioned above generally agree with each other to within a factor of $\simeq 2$, and thus give a consistent qualitative picture of cluster mass distributions and the dynamics of the cluster components. With increasing accuracy of optical and X-ray observations, however, it becomes increasingly important to assess the reliability of these methods, because any doubtless discrepancy between their results would provide important information about the internal structure and dynamics of galaxy clusters.

Gravitational lensing is sensitive to the total gravitational mass between the source and the observer irrespective of its dynamical state, and it therefore provides the most promising possibility to obtain accurate cluster masses which could then be used to infer the distribution and the support of these masses. Therefore, it appears important to quantify systematic effects which so far hamper gravitational-lensing mass estimates.

The original reconstruction technique by Kaiser & Squires was recently extended into the non-linear (strong-lensing) regime (Schneider & Seitz 1994, Seitz & Schneider 1994), and a first attempt was undertaken to remove boundary effects (Schneider 1994). It remains to be investigated whether these modifications can change reconstructed cluster masses such that they agree more closely with other methods of cluster mass determinations.

Strong lensing by galaxy clusters leads to the formation of giant luminous arcs. A mass estimate is readily found by taking the radius of a circle traced by an arc to be the Einstein radius of the cluster, and if the redshifts of lens and source are known, the mass necessary to produce an Einstein radius of the given size can be determined (for an excellent review on arcs and arclets in galaxy clusters, see Fort & Mellier 1994).

The purpose of the present paper is to demonstrate that cluster mass estimates derived from the presence of large arcs based on the assumption of spherically symmetric lenses can systematically be larger than the actual cluster mass by a factor of 1.5 to 2. This systematic deviation can be attributed to the enhanced tidal effects (shear) caused by asymmetry and substructure of the clusters. In Sect.2, it is shown that the mean surface mass density either inside a critical curve or inside a circle traced by an arc (i.e., a circle with radius equal to the distance between cluster center and arc, to which the arc is tangent) is systematically lower than the value expected from spherically symmetric models. In Sect.3, I present an analytic argument to show that deviations from spherical symmetry are generally expected to decrease the mass required for strong lensing, and in Sect.4 I discuss the results.

## 2 Mass estimates from large arcs

The numerical cluster models used for the purposes of this paper are the same as described and used in two previous papers of this series (Bartelmann & Weiss 1994, Bartelmann, Steinmetz & Weiss 1994, hereafter BSW). In short, the models were produced by $N$-body simulations starting from CDM initial conditions, normalized to the COBE quadrupole measurement of the CBR, for $\Omega_0 = 1$, $\Lambda = 0$, and $H_0 = 50$ km/s/Mpc. 13 clusters were simulated, of which 10 turned out to be critical to strong lensing at least along one direction of projection during at least one of about ten time steps per cluster between redshifts 1 and 0. Large arcs in these model clusters can be used to estimate the cluster mass, and since the true mass of the clusters is known, the accuracy of such estimates can be investigated.

Gravitational lensing is sensitive to the surface mass density $\Sigma(\boldsymbol{x})$ scaled by a redshift- (and cosmology-) dependent critical surface mass density $\Sigma_{\mathrm{cr}}$; $\boldsymbol{x}$ is an angular position vector in the lens plane. The convergence $\kappa(\boldsymbol{x})$ is defined by

$$\kappa(\boldsymbol{x}) \equiv \frac{\Sigma(\boldsymbol{x})}{\Sigma_{\mathrm{cr}}} \;, \tag{2.1}$$

with

$$\Sigma_{\mathrm{cr}} \equiv \left( \frac{4\pi G}{cH_0} \frac{r_{\mathrm{d}} r_{\mathrm{ds}}}{r_{\mathrm{s}}} \right)^{-1} \simeq 5.5 \times 10^{14} \, h \, \frac{M_\odot}{\mathrm{Mpc}^2} \, \frac{r_{\mathrm{s}}}{r_{\mathrm{d}} r_{\mathrm{ds}}} \;. \tag{2.2}$$

There, $r_{\mathrm{d,ds,s}}$ are angular-diameter distances in units of the Hubble length $c/H_0$ from the observer to the lens, from the lens to the source, and from the observer to the source, respectively, and $h$ is the Hubble constant in units of $100\,\mathrm{km/s/Mpc}$.

The lens equation provides a mapping of the lens plane onto the source plane,

$$\boldsymbol{y} = \boldsymbol{x} - \boldsymbol{\alpha}(\boldsymbol{x}) \;, \tag{2.3}$$

where $\boldsymbol{y}$ is an angular vector in the source plane, and $\boldsymbol{\alpha}(\boldsymbol{x})$ is the (appropriately scaled) deflection angle as a function of position in the lens plane. The Jacobian matrix $\mathcal{A}(\boldsymbol{x})$ of this mapping can, for a single lens plane, be written in the form

$$\mathcal{A}(\boldsymbol{x}) = \left( \frac{\partial \boldsymbol{y}}{\partial \boldsymbol{x}} \right) = \begin{pmatrix} 1 - \kappa - \gamma_1 & -\gamma_2 \\ -\gamma_2 & 1 - \kappa + \gamma_2 \end{pmatrix} \;. \tag{2.4}$$

The elements $\gamma_i$ of the trace-free part of this matrix are called shear components. Since $\mathcal{A}$ is symmetric, it has two real eigenvalues, $\lambda_\pm = 1 - \kappa \pm \gamma$, with $\gamma^2 = \gamma_1^2 + \gamma_2^2$. The curves in the lens plane defined by $\det \mathcal{A} = 0$ are called critical curves; there, one of the eigenvalues $\lambda_\pm$ vanishes.

For radially symmetric lenses, the critical curves are circles. Images close to the critical curve $\lambda_- = 1 - \kappa - \gamma = 0$ are elongated in the tangential direction to the critical curve; therefore, such critical curves are called tangential. The average of $\kappa(\boldsymbol{x})$ over the circular disk enclosed by the tangential critical curve equals unity,

$$\bar{\kappa} \equiv \frac{1}{\pi x_{\mathrm{c}}^2} \int d^2\boldsymbol{x} \, \kappa(\boldsymbol{x}) = 1 \;; \tag{2.5}$$

see, e.g., Schneider, Ehlers, & Falco (1992, henceforth SEF, Sect.8.1). $x_{\mathrm{c}}$ is the radius of the tangential critical curve.

If a large tangential arc is observed in a cluster, its distance from the center of the cluster provides an estimate for $x_{\mathrm{c}}$. By assumption of radial symmetry, $\bar{\kappa} = 1$, and therefore the cluster mass fraction contained in a cylinder of radius $x_{\mathrm{c}}$ is simply estimated to be

$$M(x_{\mathrm{c}}) = \pi x_{\mathrm{c}}^2 \, \Sigma_{\mathrm{cr}} = 1.1 \times 10^{14} \, M_\odot \left( \frac{x_{\mathrm{c}}}{30''} \right)^2 \left( \frac{r_{\mathrm{s}} r_{\mathrm{d}}}{r_{\mathrm{ds}}} \right) \;. \tag{2.6}$$

If the redshifts of the arc source and of the cluster are known, and if the cluster is radially symmetric, this mass estimate is precise. However, systematic errors arise if the radial symmetry is perturbed. Then, the critical curves are no longer circles, and the average convergence inside the critical curve with $\lambda_- = 0$ is no longer unity.

For those numerically modeled clusters which form critical curves, the convergence averaged over the area enclosed by the critical curve is readily determined. The critical curves, however, are not observed. Rather, observations provide the distance of large arcs to the cluster center. If, for example, the critical curve of a cluster is elliptic, but is assumed to be a circle with radius determined by the distance of the arc to the cluster center, the estimate of Eq.(2.6) will be inaccurate. As I will show below, the convergence averaged inside the critical curve will generally be less than unity. If large arcs are preferentially formed along the most elongated parts of the critical curve (i.e., where the critical curve is most distant from the cluster center), which is most frequently the case, then circles traced by the arcs contain the critical curve and hence enclose regions in the lens plane within which the averaged convergence is still lower.

## 2.1 Optical-depth weighted average

Since the numerical cluster models evolve significantly with decreasing redshift, they are investigated at several redshifts (typically 10 to 15) between the assumed source redshift of $z_s = 1$ and the observer at $z = 0$. Therefore, I start by determining a discrete table $\bar{\kappa}_{ijk}$ with separate entries for cluster model $i$, projected along direction $j$, and taken at timestep (or redshift) $k$. As mentioned above, I will consider two different ways of averaging $\kappa$ below. First, $\kappa$ will be averaged over the area enclosed by the critical curve of the cluster. In that case, the cluster has to be critical; if it is not, it is excluded from the analysis. Second, $\kappa$ will be averaged over areas enclosed by circles traced by "large" arcs, where "large" means that the arcs' length-to-width ratio ($L/W$) exceed a specified threshold $(L/W)_0$. Again, only clusters able to produce arcs exceeding $(L/W)_0$ will be analyzed.

Similar to the cross sections for large arcs determined in BSW, the discrete table $\bar{\kappa}_{ijk}$ is converted into smooth functions of redshift $\bar{\kappa}_{ij}(z)$, one for each cluster model $i$ and each projection direction $j$. This is done by second-order interpolation between the table entries in such redshift intervals where either the cluster model is critical to lensing or where it produces arcs satisfying $(L/W) \geq (L/W)_0$.

Having determined $\bar{\kappa}_{ij}(z)$, I average these functions separately over redshift. This average should be weighted with a factor reflecting the probability for the corresponding cluster to produce large arcs. There may be cluster models with extreme values for $\bar{\kappa}$, but if they are unlikely to form large arcs, they will not affect the mass estimates based on large arcs. An appropriate weighting function is the optical depth $\tau_{ij}(z_s)$ of the cluster model $(i,j)$ to produce large (long and thin) arcs. As in BSW, Eq.(4.1), the optical depth is defined by

$$\tau_{ij}(z_s) = \frac{1}{4\pi D_s^2} \int_0^{z_s} dz \left| \frac{dV(z)}{dz} \right| n_0 (1+z)^3 \sigma_{ij}(z, z_s) , \qquad (2.7)$$

i.e., it is the fraction of the sphere at redshift $z_s$ within which sources are imaged as large arcs. In Eq.(2.7), $D_s$ is the angular-diameter distance from the observer to the source, $\sigma_{ij}(z, z_s)$ is the cross section of cluster model $(i,j)$ at redshift $z$ to image sources at redshift $z_\mathrm{s}$ as large arcs, $n_0$ is the (supposedly constant) comoving number density of clusters, and $dV(z)$ is the proper volume of an infinitesimally thin spherical shell at redshift $z$. Since the simulated clusters are neither created nor destroyed between redshifts zero and $z_\mathrm{s} = 1$, the assumption of constant $n_0$ is satisfied. In an Einstein-de Sitter universe, Eq.(2.7) specializes to

$$\tau_{ij}(z_\mathrm{s}) = \frac{c}{H_0}\, n_0 \int_0^{z_\mathrm{s}} dz\, \sqrt{1+z}\, \left(\frac{D(z)}{D(z_\mathrm{s})}\right)^2 \sigma_{ij}(z, z_\mathrm{s})\,. \tag{2.8}$$

The redshift average over $\bar{\kappa}_{ij}(z)$, weighted by the optical depth, is then given by

$$\langle\bar{\kappa}\rangle_{ij} = \frac{1}{\tau_{ij}} \int dz\, \frac{d\tau_{ij}}{dz}\, \bar{\kappa}_{ij}(z)\,, \tag{2.9}$$

where the integration extends only over such redshift intervals within which either the cluster model is critical or where it produces large arcs. The results $\langle\bar{\kappa}\rangle_{ij}$ are weighted estimates for the mean convergences that would be derived from arc observations in clusters described by the numerical cluster models $(i, j)$.

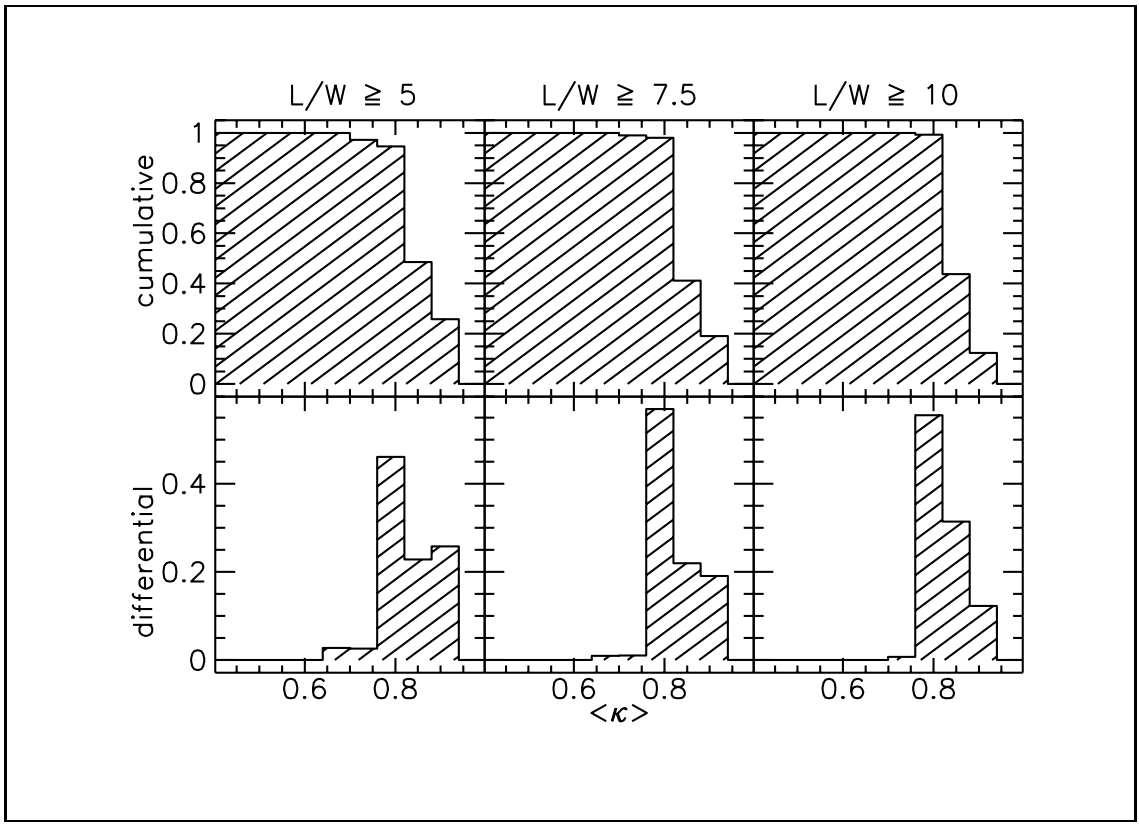

**Fig. 1.** Histograms of $\langle\bar{\kappa}\rangle_{ij}$ from averaging $\kappa$ over the interior of the tangential (outer) critical curve of the numerical cluster models. The top row shows the cumulative, the bottom row the differential distributions. The columns show the results for different length-to-width thresholds $(L/W)_0$ for the arcs, as indicated. The differential distribution shows a fairly narrow peak at around $\langle\bar{\kappa}\rangle \simeq 0.82$, quite independent of $(L/W)_0$

Finally, I determine the weighted average of $\langle\bar\kappa\rangle_{ij}$ over all cluster models. This is achieved by

$$\langle\bar\kappa\rangle = \frac{1}{\tau}\sum_{i,j}\tau_{ij}\langle\bar\kappa\rangle_{ij}\;,\qquad(2.10)$$

where $\tau=\sum\tau_{ij}$ is the total optical depth. I will focus on arcs with length-to-width ratio $(L/W)\geq(L/W)_0$ in the following, where $(L/W)_0\in\{5,7.5,10\}$ as indicated. The cross sections for such arcs were determined in, and will be taken from, BSW (see, e.g., Fig.6 there).

## 2.2 Convergence enclosed by critical curves

As mentioned before, the average convergence enclosed by the tangential critical curve of a radially symmetric lens equals unity. This quantity, however, is in general not straightforwardly related to a mass estimate because either the critical surface mass density of Eq.(2.2) or the area enclosed by the critical curve are unknown. It is however instructive to see how $\langle\bar\kappa\rangle$ deviates from unity for asymmetric cluster models. Fig.1 displays the result obtained applying Eqs.(2.8) and (2.9) to the numerical cluster models.

The upper frames in the figure display the cumulative distributions of $\langle\bar\kappa\rangle_{ij}$, the lower frames the differential distributions. The three columns of the figure distinguish between the three different thresholds $(L/W)_0$ chosen. Evidently, $\langle\bar\kappa\rangle$ is systematically smaller than unity. I investigate in the following section why this should generally be expected. The differential distributions show fairly narrow peaks around $\langle\bar\kappa\rangle\simeq0.82$, and the result is quite insensitive to the threshold $(L/W)_0$ for the arcs' length-to-width ratio. See Tab.1 below for the averages $\langle\bar\kappa\rangle$ determined from Eq.(2.10).

## 2.3 Convergence enclosed by circles traced by arcs

If the convergence is averaged within circles traced by large arcs rather than within the tangential critical curves, the distributions of $\langle\bar\kappa\rangle_{ij}$ change as displayed in Fig.2.

**Table 1.** Averaged convergence inside tangential critical curves and circles traced by large arcs, for three different thresholds in the length-to-width ratio $(L/W)_0$

|  | $(L/W)_0$ | | |
|---|---|---|---|
| $\langle\bar\kappa\rangle$ | 5 | 7.5 | 10 |
| enclosed by tangential critical curve | 0.83 | 0.82 | 0.82 |
| enclosed by circles traced by arcs | 0.64 | 0.64 | 0.67 |

As expected, the distributions shift towards lower averaged convergence. Table 1 summarizes the averages $\langle\bar\kappa\rangle$ for the three length-to-width thresholds chosen and for averages performed within the tangential critical curves or within circles traced by arcs. Again, the results are fairly insensitive to the length-to-width threshold specified. The distributions shown in Fig.2 are broader than those of Fig.1 because of the scatter in the positions of large arcs relative to the cluster center.

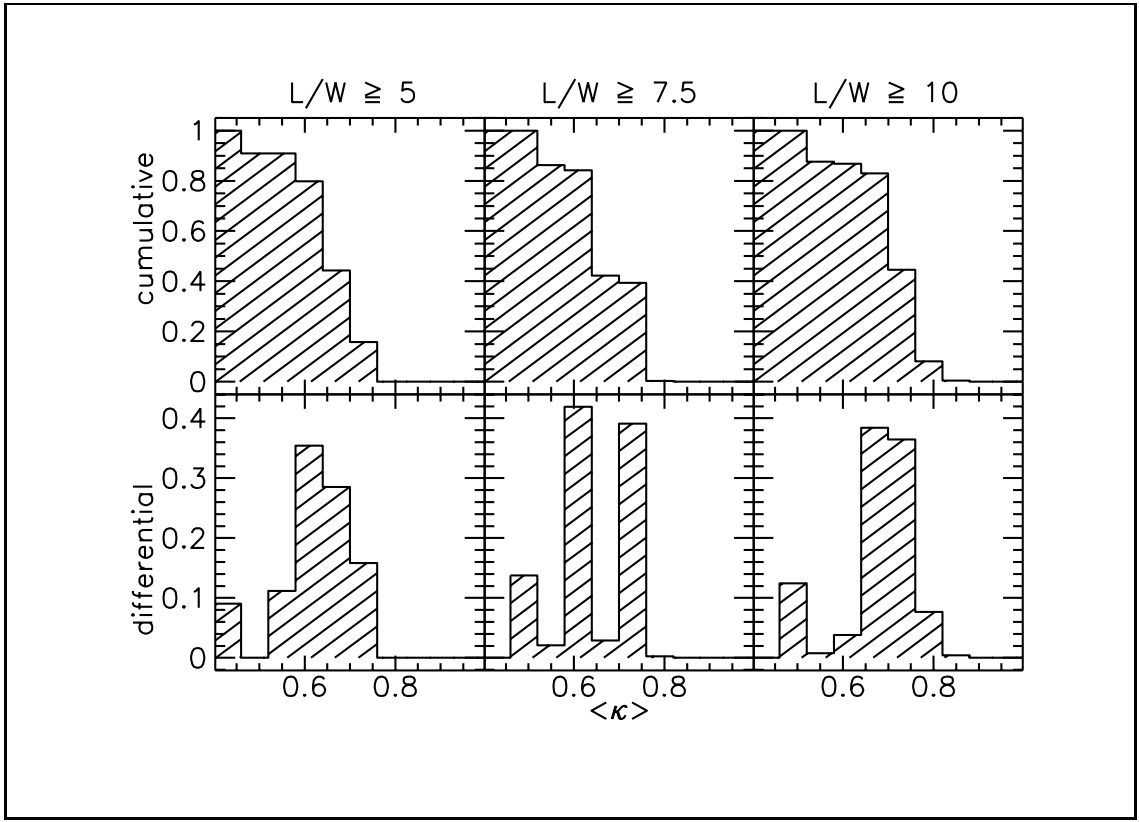

**Fig. 2.** Histograms of $\langle\bar\kappa\rangle_{ij}$ from averaging $\kappa$ over circles centered on the cluster center and traced by large arcs. As in Fig.1, the top row shows the cumulative, the bottom row the differential distributions. The columns show the results for different length-to-width thresholds $(L/W)_0$ for the arcs, as indicated. The distribution is significantly broader than before, and the average is markedly smaller than unity

## 3 Analytic considerations

The results presented in Sect.2 and in Figs.(1,2) suggest that for asymmetric lens models the mean convergence within the critical curve defined by $\lambda_- = 1 - \kappa - \gamma = 0$ is generally less than unity, $\bar\kappa < 1$; i.e., that it is smaller than the value expected for radially symmetric lenses. The following analytic consideration shows why this should be expected.

Let the tangential critical curve be parametrized by $\boldsymbol{c}(\varphi)$, where $\varphi$ is a continuous parameter. Critical curves are closed, and therefore they define an interior domain $C \subset \mathbb{R}^2$. If the area of this domain is $A_c$, the convergence averaged over $C$ is given by

$$\bar\kappa_c = \frac{1}{A_c} \int d^2\boldsymbol{x}\, \kappa(\boldsymbol{x})\,, \tag{3.1}$$

with

$$A_c = \int d^2\boldsymbol{x}\,. \tag{3.2}$$

Using the identity $\nabla \cdot \boldsymbol{x} = 2$ in two dimensions, and applying Gauss' theorem in the plane, Eq.(3.2) can be written

$$A_c = \frac{1}{2} \oint d\varphi\, (\boldsymbol{c} \cdot \boldsymbol{n})\,, \tag{3.3}$$

where $\boldsymbol{n}$ is the outward-directed vector orthogonal to the critical curve at the point $\boldsymbol{c}(\varphi)$. This orthogonal vector can be written

$$\boldsymbol{n} = \mathcal{R}\left(-\frac{\pi}{2}\right) \cdot \boldsymbol{t} = \begin{pmatrix} \dot{c}_2 \\ -\dot{c}_1 \end{pmatrix} . \tag{3.4}$$

There, $\boldsymbol{t}$ is the tangent vector to the critical curve, and $\mathcal{R}$ is the two-dimensional rotation matrix. The dots symbolize the derivative with respect to the curve parameter $\varphi$.

Because of Poisson's equation, the convergence $\kappa$ satisfies

$$\kappa = \frac{1}{2} \, \nabla \cdot \boldsymbol{\alpha} \,, \tag{3.5}$$

where $\boldsymbol{\alpha}$ is the deflection angle (SEF, Sect.5.1). Therefore, Gauß' theorem can again be applied to write

$$\int d^2\boldsymbol{x} \, \kappa = \frac{1}{2} \oint d\varphi \, (\boldsymbol{\alpha} \cdot \boldsymbol{n}) \,. \tag{3.6}$$

With the coordinate representation (3.4) of $\boldsymbol{n}$, the integral in (3.6) can be transformed to

$$\oint d\varphi \, (\boldsymbol{\alpha} \cdot \boldsymbol{n}) = \oint d\varphi \, (\alpha_1 \dot{c}_2 - \alpha_2 \dot{c}_1) = \oint d\varphi \, (\dot{\alpha}_2 c_1 - \dot{\alpha}_1 c_2)$$
$$= \oint d\varphi \, \left[\boldsymbol{c} \cdot \mathcal{R}\left(-\frac{\pi}{2}\right) \cdot \dot{\boldsymbol{\alpha}}\right] , \tag{3.7}$$

where I have integrated by parts and used that the integral is performed along a closed loop. The lens equation (2.3) can now be used to replace $\dot{\boldsymbol{\alpha}}$ by $(\boldsymbol{t} - \mathcal{A} \cdot \boldsymbol{t})$, where $\mathcal{A}$ is the Jacobian matrix of the lens mapping. Substituting this together with (3.7) into (3.6), we obtain

$$\bar{\kappa}_\mathrm{c} = 1 - \frac{1}{2 A_\mathrm{c}} \oint d\varphi \, \left[\boldsymbol{c} \cdot \mathcal{R}\left(-\frac{\pi}{2}\right) \mathcal{A} \mathcal{R}\left(\frac{\pi}{2}\right) \cdot \boldsymbol{n}\right] , \tag{3.8}$$

where I have used Eqs.(3.3) and (3.4).

For radially symmetric lenses, the tangential vector to the (circular) tangential critical curve is contained in the kernel of $\mathcal{A}$, and therefore $\mathcal{A}\mathcal{R}(\pi/2) \cdot \boldsymbol{n}$ vanishes along the tangential critical curve. Then, we recover from (3.11) the result that $\bar{\kappa} = 1$ for radially symmetric lenses. In the generic case, however, $\boldsymbol{t}$ is contained in the kernel of $\mathcal{A}$ only at cusps, and $\mathcal{A}$ is a positive semi-definite matrix along the tangential critical curve. Then, also $\mathcal{R}(-\pi/2)\mathcal{A}\mathcal{R}(\pi/2)$ is positive semi-definite, and therefore

$$\boldsymbol{n} \cdot \mathcal{R}\left(-\frac{\pi}{2}\right) \mathcal{A} \mathcal{R}\left(\frac{\pi}{2}\right) \cdot \boldsymbol{n} \geq 0 \,. \tag{3.12}$$

Thus, the angle between the image of the normal vector to the critical curve, which is normal to the caustic, and the normal vector itself, is less than or equal to $\pi/2$. If the lens does not deviate too strongly from radial symmetry, then the coordinate origin can be chosen such that $\boldsymbol{c}$ has approximately the same direction as $\boldsymbol{n}$, and then the line integral in (3.8) will be non-negative because of the positive semi-definiteness of the Jacobian matrix $\mathcal{A}$. Therefore, we expect from Eq.(3.8) that the average convergence enclosed by the tangential critical curve will generally be less than unity if the lens is not radially symmetric.

In particular, consider a circle enclosing the tangential critical curve. If $\kappa$ is decreasing outward from the lens center, $\mathcal{A}$ is positive semi-definite along that circle. If the center of the circle is chosen as the coordinate origin, $\boldsymbol{r} = r(\cos\varphi, \sin\varphi)$ if $r$ is the radius of the circle, and therefore $\boldsymbol{r} = \boldsymbol{n}$. Then, Eq.(3.8) can be written

$$\bar{\kappa} = 1 - \frac{1}{2A_\mathrm{c}} \oint d\varphi \left[ \boldsymbol{n} \cdot \mathcal{R}\left(-\frac{\pi}{2}\right) \mathcal{A} \mathcal{R}\left(\frac{\pi}{2}\right) \cdot \boldsymbol{n} \right] . \qquad (3.9)$$

Because of the positive semi-definiteness of $\mathcal{A}$ along the circle, the line integral is non-negative, and thus $\bar{\kappa} \le 1$ averaged inside the circle. This explains why cluster mass estimates assuming spherical symmetry are systematically too high. Note that Subramanian & Cowling (1986) have already noted that $\bar{\kappa}$ can be less than unity for lenses with elliptical symmetry, and that they have given an example for a (somewhat unrealistic) lens model with arbitrarily low $\bar{\kappa}$.

## 4 Summary and conclusions

Using the numerical cluster models described by BSW, I have investigated how much mass there is required for producing large (i.e., long and thin) arcs with asymmetric lenses. The quantity which is naturally used for addressing this question is the surface mass density of the lens in units of its critical value, that is, the convergence $\kappa$, averaged over areas in the lens plane either enclosed by the tangential critical curve or by circles traced by large arcs. For radially symmetric lenses, this quantity is unity, and it therefore provides a simple estimate for cluster masses required for large arcs once the redshifts of arc sources and cluster lenses are known.

The average convergence enclosed by the tangential critical curve is instructive to investigate, but it is not observable since the critical curve is unknown. Rather, the distances of large arcs from the cluster centers provide an estimate for the Einstein radius of an equivalent radially symmetric lens. If, according to the assumption of radial symmetry, the average convergence is taken to be unity inside the so-estimated Einstein radii, systematic errors in the cluster masses result.

Using the numerical cluster models, I have shown that the average convergence enclosed by the tangential critical curves is systematically smaller than unity, $\langle\bar{\kappa}\rangle \simeq 0.82$. This result was achieved weighting the values of $\bar{\kappa}$ obtained from the individual cluster models with the optical depths of these clusters for producing large arcs. An analytic investigation was presented to demonstrate why perturbations to the radial symmetry of lenses should in general result in lowering $\bar{\kappa}$. The physical reason behind this is that part of the convergence necessary for producing large arcs in spherically symmetric models can be replaced by the enhanced tidal effects (shear) resulting from asymmetries and substructures in aspheric lenses.

Taking the average over the convergence inside circles traced by large arcs further reduces the mass estimate; then, $\langle\bar{\kappa}\rangle \simeq 0.64$. If this value is compared to the mass estimate based on the assumption of radial symmetry, one can conclude from Tab.1 and Fig.2 that on average the cluster mass required for large arcs will be lower by a factor of $\simeq 1.6$ than expected from radially symmetric models, and that there is a probability of $\simeq 20\%$ for overestimating the actual cluster mass by a factor of $\simeq 2$. Interestingly, a recent analysis of the cluster A 1689 (Daines, Jones, & Forman 1994) argues that the X-ray mass of this cluster is lower by a factor of $\simeq 2$ than the mass required for the arcs observed, if the lens is taken to be radially symmetric.

In view of this systematic overestimate of cluster masses from the strong lensing effect, doubts may be raised whether the non-thermal pressure support of the intracluster gas suggested by Miralda-Escudé & Babul (1994) and Loeb & Mao (1994) is indeed required from a comparison of X-ray mass estimates to lensing mass estimates in several galaxy clusters. A moderate deviation of $\bar{\kappa} = 0.85$ from unity was assumed in the analysis of the cluster A 2218, and this reduction was motivated by the obvious asymmetry of the cluster and quantified by a lens model which reproduces the gross features of the arcs in that cluster. However, even $\bar{\kappa} = 0.85$ may still overestimate the lensing mass of the cluster by a factor of $\simeq 1.3$. Since projection effects may further decrease the claimed discrepancy between the masses required for lensing and for the X-ray emission, it appears as if the conclusion about a non-thermal pressure support of the intracluster gas could be avoided. Note that this does not mean that there is no non-thermal pressure support, but it means that strong gravitational lensing does not unambiguously *require* it. The systematic effect discussed in this paper should add a cautionary remark to comparisons of cluster masses derived from strong gravitational lensing with results required for the X-ray emission.

Finally, there is a puzzling discrepancy between the result found here, that cluster masses are likely to be *lower* than estimated from strong lensing by factors of $1.6\ldots2$, and the results from weak-lensing cluster reconstructions (Fahlman et al. 1994, Carlberg et al. 1994), that weak lensing in the cluster MS 1224+20 requires a cluster mass $2\ldots3$ times *higher* than the virial mass. The example of the cluster A 2169, where the weak-lensing reconstruction yields a mass distribution compatible with zero mass although the cluster is the hottest and most luminous X-ray cluster in the sky, shows that there may also be systematic uncertainties in the reconstructed cluster masses which are not yet understood.

*Acknowledgements.* It is a pleasure to thank Matthias Steinmetz and Achim Weiss for their substantial contribution to the project to which this study belongs. I am also grateful to Martin Hähnelt, Chris Kochanek, Abraham Loeb, Jordi Miralda-Escudé, Ramesh Narayan and Peter Schneider for comments and discussions.